\documentclass[aps,prc,twocolumn,showpacs,superscriptaddress, twoside,showpacs]{revtex4-2}

\usepackage{graphicx}
\usepackage{dcolumn}
\usepackage{bm}
\usepackage{booktabs}
\usepackage{amsmath}
\usepackage{color}
\usepackage{makecell}
\usepackage{tabularx}
\usepackage{multirow}

\begin{document}
\preprint{APS/123-QED}

\title{A high-lying isomer in $^{92}$Zr with lifetime modulated by the atomic charge states: a proposed approach for a nuclear gamma-ray laser}
		\author{\mbox{C. X. Jia }}
        \author{\mbox{S. Guo}}	
        \email{gs@impcas.ac.cn}
        \author{\mbox{B. Ding }}
        \email{dbing@impcas.ac.cn}
        \author{\mbox{X. H. Zhou}}
        \affiliation{State Key Laboratory of Heavy Ion Science and Technology, Institute of Modern Physics, Chinese Academy of Sciences, Lanzhou 730000, China}
        \affiliation{School of Nuclear Science and Technology, University of Chinese Academy of Science, Beijing 100049, China}

        \author{\mbox{C.X. Yuan}}
        \email{yuancx@mail.sysu.edu.cn}
        \author{\mbox{W. Hua }}
        \affiliation{Sino-French Institute of Nuclear Engineering and Technology, Sun Yat-sen University,  Zhuhai, 519082, Guangdong, China}
        
	  \author{\mbox{J. G. Wang}}	
        \author{\mbox{S. W. Xu}}

        \affiliation{State Key Laboratory of Heavy Ion Science and Technology, Institute of Modern Physics, Chinese Academy of Sciences, Lanzhou 730000, China} 
		\affiliation{School of Nuclear Science and Technology, University of Chinese Academy of Science, Beijing 100049, China}
  
        \author{\mbox{C. M. Petrache}}
		\affiliation{University Paris-Saclay, CNRS/IN2P3, IJCLab, Orsay, 91405, France}
		\author{\mbox{E. A. Lawrie}}
		\affiliation{iThemba LABS, National Research Foundation, PO Box 722, 7131 Somerset West, South Africa}
        \author{\mbox{Y. B. Wu}}
		\affiliation{Max-Planck-Institut f$\ddot{u}$r Kernphysik, Saupfercheckweg 1, D-69117 Heidelberg, Germany}
        
        \author{\mbox{Y. D. Fang}}
        \author{\mbox{Y. H. Qiang}}
        \author{\mbox{Y. Y. Yang}}
        \author{\mbox{J. B. Ma}}
        \author{\mbox{J. L. Chen}}
        \author{\mbox{H. X. Chen}}
	  \author{\mbox{F. Fang}}
	  \author{\mbox{Y. H. Yu}}
	  \author{\mbox{B.F. Lv}}
	  \author{\mbox{F. F. Zeng}}
	  \author{\mbox{Q. B. Zeng}}
	  \author{\mbox{H. Huang}}
	  \author{\mbox{Z. H. Jia}}
        \author{\mbox{W. Liang}}
        \author{\mbox{W. Q. Zhang}}
	  \author{\mbox{J. H. Li}}
	  \author{\mbox{J. H. Xu}}
        \author{\mbox{M. Y. Liu}}
	  \author{\mbox{Y. Zheng}}
        \author{\mbox{Z. Bai}}
	  \author{\mbox{S. L. Jin}}
	  \author{\mbox{K. Wang}}
        \author{\mbox{F. F. Duan}}
	  \author{\mbox{G. Yang}}  
        \author{\mbox{G. S. Li}}
  	\author{\mbox{M. L. Liu}}
        \author{\mbox{Z. Liu}}
        \author{\mbox{Z. G. Gan}}
        \author{\mbox{M. Wang}}
        \author{\mbox{Y. H. Zhang}}
        \affiliation{State Key Laboratory of Heavy Ion Science and Technology, Institute of Modern Physics, Chinese Academy of Sciences, Lanzhou 730000, China} 
		\affiliation{School of Nuclear Science and Technology, University of Chinese Academy of Science, Beijing 100049, China}

        \author{\mbox{Y. Q. Liang}}
        \affiliation{Sino-French Institute of Nuclear Engineering and Technology, Sun Yat-sen University, Guangdong Zhuhai 519082, China}

        \author{\mbox{Wei Rui}}
        \author{\mbox{S. Q. Li}}
        \affiliation{Department of Physics, Guangxi Normal University, Guilin 541004, Guangxi, China}

		\affiliation{Department of Physics $\&$ Astronomy, University of the Western Cape, P/B X17, Bellville ZA-7535, South Africa}
  
		\author{\mbox{H. J. Ong}}
		\affiliation{State Key Laboratory of Heavy Ion Science and Technology, Institute of Modern Physics, Chinese Academy of Sciences, Lanzhou 730000, China} 
		\affiliation{School of Nuclear Science and Technology, University of Chinese Academy of Science, Beijing 100049, China}
		\affiliation{Joint Department for Nuclear Physics, Lanzhou University and Institute of Modern Physics, Chinese Academy of Sciences, Lanzhou 730000, China}
		\affiliation{Research Center for Nuclear Physics, Osaka University, Osaka 567-0047, Japan}
		
		\author{\mbox{Y. Li}}
		\affiliation{Hebei University, Baoding 071001, People's Republic of China}
		\author{\mbox{N. W. Huang}}
		\author{\mbox{L. J. Liu}}
		\affiliation{Department of Physics, Huzhou University, Huzhou 313000, China}

		\author{\mbox{A. Rohilla}}
		\affiliation{State Key Laboratory of Heavy Ion Science and Technology, Institute of Modern Physics, Chinese Academy of Sciences, Lanzhou 730000, China}

 

\date{\today}

\begin{abstract}
The nuclides $^{92}$Zr are produced and transported by using a radioactive beam line to a low-background detection station. After a flight time of about 1.14~$\mu$s, the ions are implanted into a carbon foil, and four $\gamma$ rays deexciting the 8$^+$ state in $^{92}$Zr are observed in coincidence with the implantation signals within a few nanoseconds. We conjecture that there exists an isomer located slightly above the 8$^+$ state in $^{92}$Zr. The isomeric lifetime in highly charged states is extended significantly due to the blocking of internal conversion decay channels, enabling its survival over the transportation. During the slowing-down process in the carbon foil, the $^{92}$Zr ions capture electron and evolve toward neutral atoms, and consequently the lifetime is restored to a normal short value. Such a high-lying isomer depopulated by a low-energy transition may provide unique opportunity to develop nuclear ${\gamma}$ laser.
\end{abstract}

\maketitle
The long-sought nuclear gamma-ray laser (NGL) remains a great challenge despite numerous efforts\cite{Rivlin2007}. Tremendous endeavors focus primarily on the development of two conflicting critical techniques: achieving sufficient accumulation of nuclei in the upper laser level and narrowing the gamma-ray linewidth emitted to its natural radiative width\cite{Rivlin2007,RMP691085,PRR7023229,PRL106162501}.

\begin{figure}[ht]
    \centering
    \includegraphics[scale=0.25]{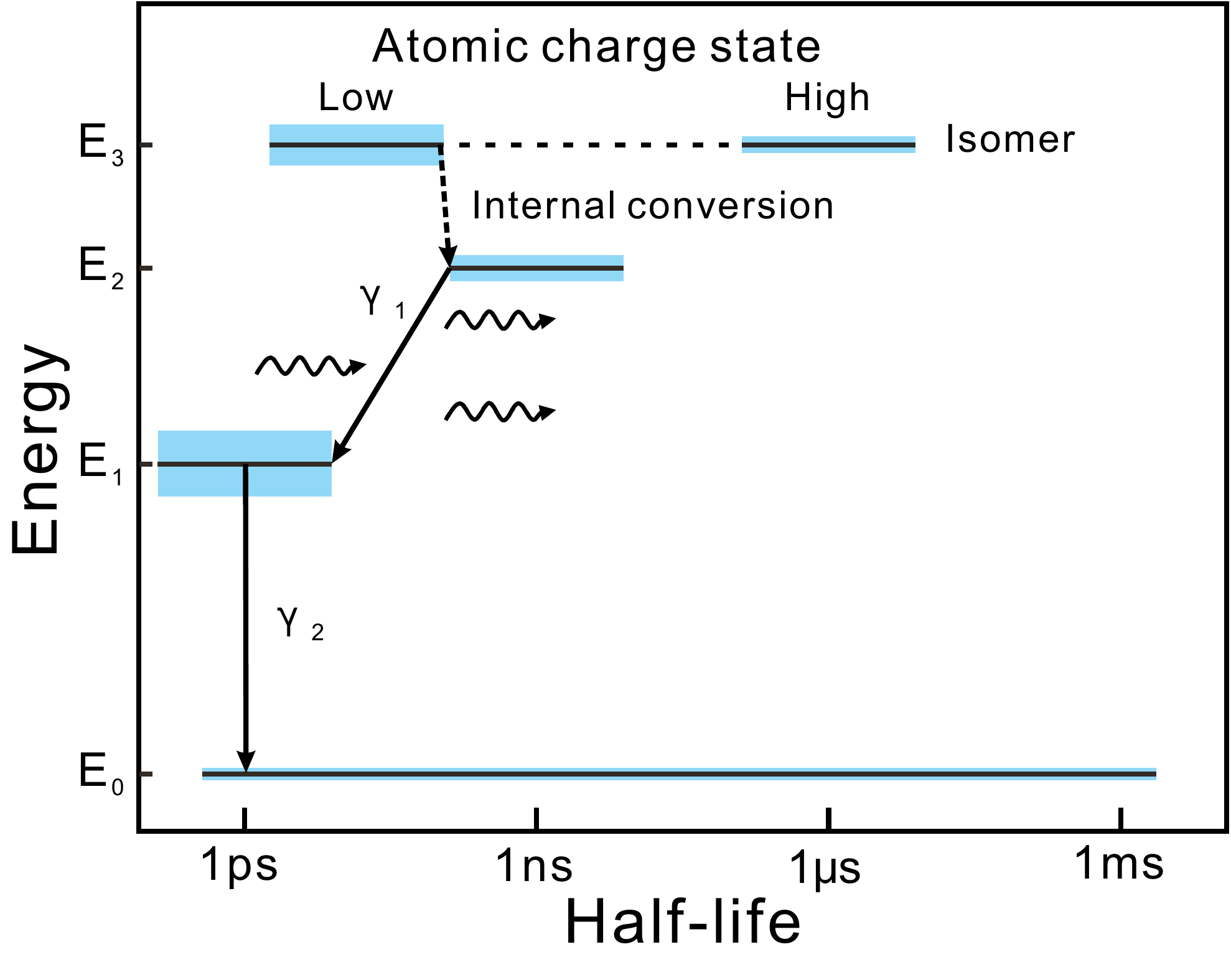}
    \caption{The diagram illustrates an ideal four-level nuclear laser scheme. Blue shading indicates the spectral linewidth. The ions can be stored in the isomeric state E$_3$ with a relatively long lifetime in highly-charged atomic charge states, and promptly decay to the upper laser level E$_2$ via changing the charge state. The lower laser level E$_1$ has a short lifetime to achieve a sufficiently broad linewidth of $\gamma_1$. The E$_0$ level is a lower-lying excitation state or the ground state.}
    \label{4level}
\end{figure}

To date, most of the pump methods are based on two elementary processes\cite{Rivlin2007}: the resonance absorption of X-ray photons and the radiative capture of neutrons. Here we propose a new pump method via a special nuclear metastable state, namely isomer\cite{10.1063/1.1996473}, to meet the two requirements simultaneously. Such an isomer has a high excitation energy and decays by a cascade of gamma transitions, as shown in Fig. \ref{4level}. The states in the decay paths form a structure partially similar to the standard four-level scheme of visible lasers\cite{2003Laser}. Supposing that the isomers in highly-charged states are produced with a large quantity and long-lived due to blocking of internal conversion decay channels, we can change the charge states and hence the isomeric lifetime\cite{PhysRevLett.62.1025,ROSE196615,PhysRevLett.75.1715,PhysRevC.91.031301}.  If the isomer in neutral atom has very short lifetime and decays promptly, we can realize a particle number inversion between the upper and lower laser levels. The lifetime of the lower laser level is significantly shorter than that of the upper level, and therefore the population inversion can be ensured by the promptly decay of the lower level. In the meantime, the short lifetime of the lower laser level would lead to a relatively broad natural radiative width. The isomeric beam ions can be cooled down and controlled, which in principle narrows the emission linewidth by eliminating Doppler broadening.

\par

To realize this method, an appropriate isomer is of critical importance. However, high-lying isomers depopulated by low-energy transitions are so far scarcely known, partly due to the difficulty from the experimental side. In this letter, we report the identification of an isomer in $^{92}$Zr as a potential candidate for NGL.

The experiment was carried out at the Heavy Ion Research Facility in Lanzhou (HIRFL), China. Fusion-evaporation residues were produced using a 559 MeV $^{86}$Kr beam bombarding $^{12}$C targets, and transported through the Radioactive Ion Beam Line in Lanzhou (RIBLL)\cite{SUN2003496} to a low $\gamma$ ray background detection station. Two 100 $\mu$g/cm$^2$ carbon foils were installed on both sides of the target frame at the primary target position of RIBLL. The magnetic fields of RIBLL were set to select the fusion-evaporation residues, while the primary beam was deflected away. After about 1.14 $\mu$s flight, the ions were implanted into a 7~cm × 7~cm × 3~mm plastic scintillator detector covered with a 20~$\mu$m carbon foil. The implantation rate in the plastic detector was around 61~kHz.

Subsequent gamma rays emitted from decays and secondary reactions were detected by five high-purity germanium (HPGe) detectors equipped with anti-Compton shields. Among them, four coaxial type detectors were installed perpendicular to the beam direction, while one segmented clover detector faces it. Experimental details can be found in Ref.\cite{PhysRevLett.128.242502}.\par
 
At the detection station, most of the $\gamma$ rays originate from the environmental background and the decay of the isomeric states of the implanted ions. The ground states of the implanted ions have quite long lifetimes, and hence the $\gamma$ rays from their $\beta$ decay are invisible. 
\begin{figure*}[ht]
    \centering
    \includegraphics[scale=0.8]{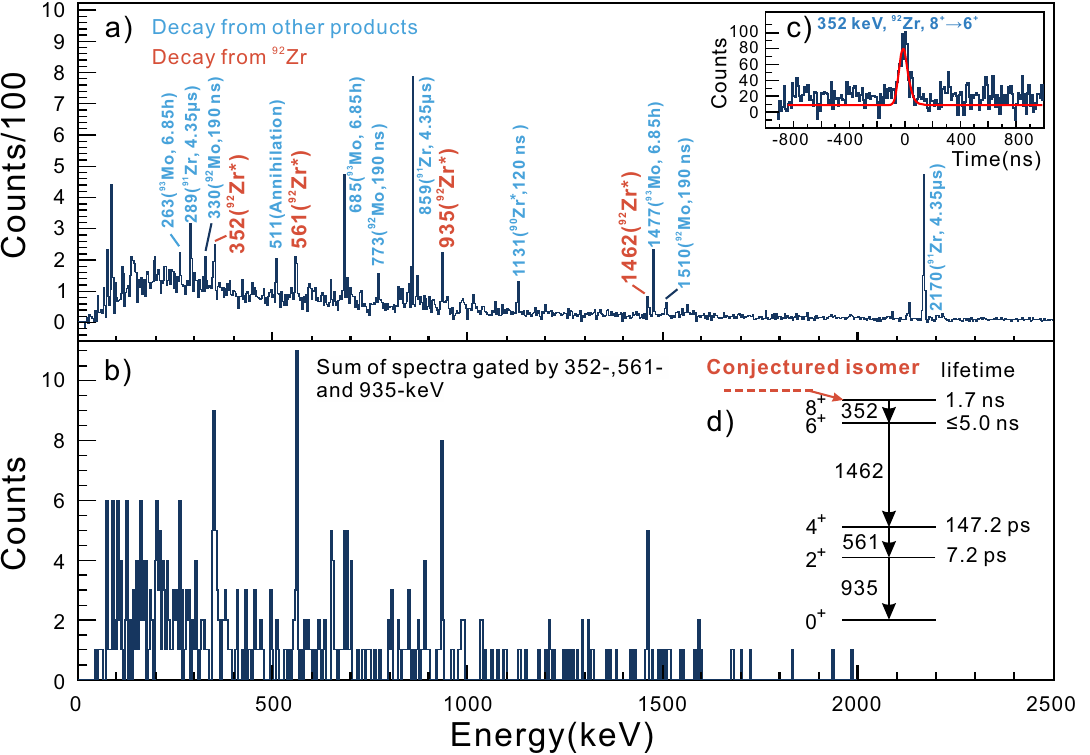}
    \caption{(a) Implantation-coincident $\gamma$-spectrum with background subtracted. (b) Gated spectrum demonstrating the coincidences for the 935-, 561-, 1462-, and 352-keV transitions. (c)The time spectra between the gamma rays and implantation signals recorded by the HPGe detector and scintillator. (d) Partial level scheme of $^{92}$Zr together with the conjectured isomeric state.}
    \label{Energy}
\end{figure*}\par

Fig.\ref{Energy}(a) shows the measured $\gamma$-ray spectrum in coincidence with the implantation events recorded by the plastic scintillator, for which the background obtained in the pre-implantation time window was properly subtracted. In this figure, $\gamma$ rays from long-lived isomers and the background are effectively suppressed, and in contrast, those from short-lived isomers become prominent. Most of the observed $\gamma$ peaks in the spectrum were assigned to known isomers in $^{92,93}$Mo and $^{90,91}$Zr. Interestingly, the 352-, 1462-, 561- and 935-keV transitions are evident, but not belonging to any known isomeric de-excitation in this mass region. These transitions exhibit energies identical to the first four transitions (8$^+$ $\rightarrow $6$^+$, 6$^+$ $\rightarrow $4$^+$, 4$^+$ $\rightarrow $2$^+$, 2$^+$ $\rightarrow $0$^+$) in $^{92}$Zr, as shown in Fig.\ref{Energy}(d)~\cite{PhysRevC.96.024314}. We checked the $\gamma$-$\gamma$ coincidence relationship correlated with the implantation signals. The gated spectrum is presented in Fig.\ref{Energy}(b). Although the statistics are low, the prompt coincidences among them are obvious. Furthermore, we found that the intensities of the four transitions are almost same after correction for the detection efficiencies. Therefore, it is concluded that the four $\gamma$ rays are emitted sequentially from the 8$^+$ state of $^{92}$Zr. The lifetime of the 8$^+$ state in $^{92}$Zr was reported to be only 1.7(10)~ns\cite{BAGLIN20122187}, and it could not survive the flight over RIBLL. Therefore, we propose that the 8+ state is fed by the decay of an isomer nearby. 

The kinetic energies of the ions are about 5 A MeV, which enable secondary fusion-evaporation reactions or Coulomb excitation of nuclei in the carbon foil. Coulomb excitation can simply be excluded; it would preferentially populate low-lying states, leading to a much stronger 935 keV peak compared with other transitions. For the secondary fusion-evaporation reactions, we examined the reactions between the major ions and carbon, and no $\gamma$ rays are observed from the secondary reaction products. As $^{92}$Zr is merely a minor by product in secondary reactions, its detectable yield originates predominantly from primary reaction products. However, the four $\gamma$ rays emitted almost simultaneously with the implantation events. It is puzzling that the lifetime of the proposed isomer should be long enough to survive in flight, on the other hand, it should be quite short for the prompt decay after implantation.\par

The only reasonable explanation is that the lifetime of this isomer is changeable. This is possible because changes in charge states can lead to a dramatic change in lifetime. Most of the $^{92}$Zr ions have a charge state between 30$^+$ and 36$^+$ in flight and the electrons fill the K shell and partial orbits of the L shell. If the transition energy between the proposed isomer and the 8$^+$ state is smaller than the electron binding energies of the ocuupied orbits, the internal conversion channels are completely blocked and the lifetime of this state is dramatically extended\cite{PhysRevC.96.031303,PhysRevC.91.031301,PhysRevC.89.044308}. Once the highly charged ions pass through the carbon foil, they capture electrons during the slowing-down process and evolve toward neutral atoms. The isomeric lifetime recovers to a normal short value, causing it to decay rapidly. In such a manner, the 935-, 561-, 1462- and 352-keV $\gamma$ rays are emitted in a few nanoseconds after implantation. We note that the transition linking the inferred isomer to 8$^+$ was not observed in the present work.

The isomer has different lifetimes during flight over the beam line and stopping in the solid material, denoted by $\tau_f$ and $\tau_s$ , respectively. $\tau_s$ can be deduced from the analysis of the time spectrum between the $\gamma$ rays and implantation signals measured by the HPGe detectors and the plastic scintillator. By fitting the time spectrum\cite{PhysRevC.94.044316,MACH198949}, $\tau_s$ is extracted as 1 ns, with a 3$\sigma$ uncertainty of 38 ns, as shown in Fig. \ref{Energy}(c). Consequently, the upper limit of $\tau_s$ is set to 39 ns. $\tau_f$ can be estimated from the loss of isomeric ions during flight\cite{TARASOV20084657,Bethe,RN20,TARASOV,RN21}. The flight lifetime of $^{92m}$Zr is derived as $\tau_f > 853$~ns.

\begin{figure}[ht]
    \centering
    \includegraphics[scale=0.6]{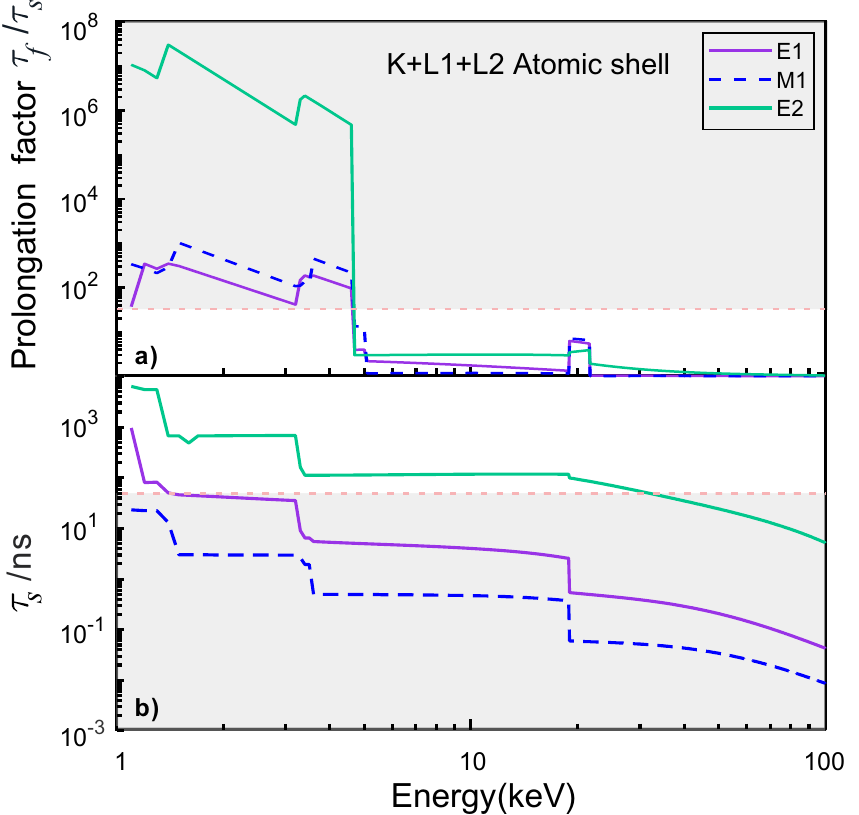}
    \vspace{-1em}
    \caption{The prolongation factor and the $\tau_s$ changes with the energy of the transition depopulating the isomer, assuming E1, M1 and E2 character, respectively. }
    \label{untitled}
\end{figure}

Both $\tau_s$ and $\tau_f$ are related to the electromagnetic multipolarity and the energy of the transition that deexcites this isomer. Based on the extracted limits of the lifetimes of $^{92m}$Zr during flight and slowing-down processes, we investigated whether there exist specific transitions that are in line with the extracted lifetimes.

For $^{92m}$Zr ions in flight, the prolongation factor is defined as:
\begin{equation}
    \frac{\tau_f}{\tau_s}=\frac{1+\alpha_{total}}{1+\alpha_Q},
\end{equation}where $\alpha_{total}$ is the total internal conversion coefficient in a neutral atom and $\alpha_Q$ is the internal conversion coefficient of an ion in the charge state of $Q$. The relationship between the internal conversion coefficient and the transition energy was calculated via $BrIcc$~\cite{Kibdi2008EvaluationOT}, with the ionization energy correction for the charge states also considered \cite{NIST_ASD}. From the extracted upper limit of $\tau_s$ and the lower limit of $\tau_f$, the prolongation factor is deduced to be greater than 22, as shown in Fig. \ref{untitled}a. 

For the stopped $^{92m}$Zr ions, $\tau_s$ = $1/[\lambda_\gamma(1+\alpha_{total})]$. Considering that the transition energy should be low and the lifetime is not longer than a few nanoseconds, only the E1, E2, and M1 transitions are permitted. In $A=45\sim150$ mass region the maximum $\gamma$ transition strengths for the E1, E2, and M1 transitions were reported as 0.01, 300, and 3 W.u. in Ref.\cite{2013viii}, respectively. Using the Weisskopf estimate and the internal conversion coefficient for a given transition energy and type, we derive the energy-$\tau_s$ relationship as shown in Fig. \ref{untitled}b. $\tau_s$ should be smaller than the deduced upper limit of 39 ns.

Combining both the prolongation factor and $\tau_s$, the allowed transition energy ranges for each multipolarity type were deduced (Figure \ref{untitled}). E2 transitions are excluded, while E1 and M1 transitions are possible. The transition energy ranges are 3.2-5 keV for E1 and <5 keV for M1. The corresponding possible spin-parities are 9$^+$, 9$^-$, 8$^+$, 8$^-$, 7$^+$, and 7$^-$. According to the selection rule, 8$^+$, 7$^+$ and 7$^-$ can link to the 6$^+$ state by M1/E2 or E1 transition with much higher transition energy, but such a transition is not observed. Therefore, these spin-parity values can be ruled out.

\begin{figure}[ht]
    \centering
    \includegraphics[scale=0.83]{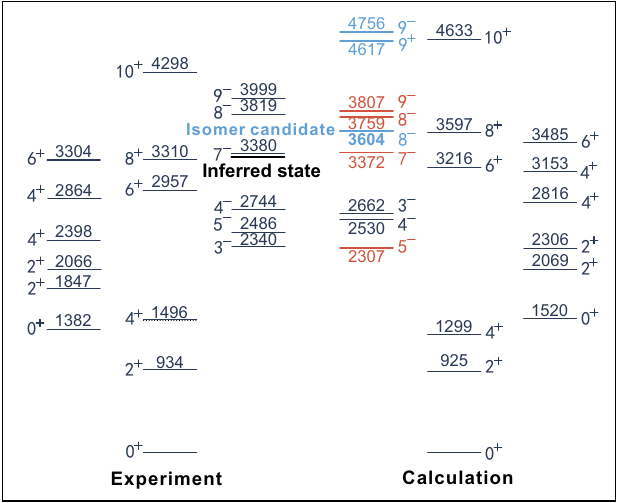}
    \vspace{-1em}
    \caption{The diagram shows experimental and calculated states of $^{92}$Zr. The black line is the inferred state, the red line indicates the calculated negative parity state and the blue line represents the candidate levels for the proposed isomer.}
    \label{levelscheme}
\end{figure}

The above discussion suggests a charge-state-dependent isomeric state in $^{92}$Zr located slightly above the 8$^+$ state and with constrained spin-parity values. It is crucial to verify whether such a state might exist theoretically. Shell-model calculations for near-spherical $^{92}$Zr was performed using a Hamiltonian constructed from the JUN45 interaction\cite{PhysRevC.80.064323,PhysRevLett.104.012501}, taking $^{56}$Ni as a doubly closed core. The model space consists of the $\pi$($0f_{5/2}$, $1p_{3/2}$, $1p_{1/2}$ and $0g_{9/2}$) and $\nu$($0f_{5/2}$, $1p_{3/2}$, $1p_{1/2}$, $0g_{9/2}$, $0g_{7/2}$, $1d_{5/2}$, $1d_{3/2}$, $2s_{1/2}$, and $0h_{11/2}$)single-particle orbitals. Fig.\ref{levelscheme} shows the experimental and calculated low-lying states in $^{92}$Zr. Generally, the theoretical calculations reproduce the states well. Among the calculated $9^+$, $9^-$ and $8^-$ states, the lowest one is the first $8^-$ state with a calculated excitation energy of 3604 keV. Notably, the calculated energies between this $8^-$ state and the first $8^+$ state is only a few keV. In the previous experimental work\cite{PhysRevC.89.044308}, $7^-$, $8^-$, and $9^-$ states were reported, with linking transitions among them. They are expected to have similar configurations. These experimental states correspond to the $7_1^-$, $8_2^-$, and $9_1^-$ states predicted theoretically, respectively. Compared to them, the calculated $8_1^-$ state has a larger $\nu h_{11/2}$ component. The occupation of a quasiparticle in the $h_{11/2}$ orbital may drive the core to a larger deformation with a decrease of the excitation energy. The main quasiparticle configuration of the calculated first $8^-$ state consists of $\pi p_{3/2}g_{9/2}$ and $\nu h_{11/2}d_{5/2}$ components. The E1 transition linking it to the $8^+$ state can be enhanced by the octupole correlations between proton $p_{3/2}$ and $g_{9/2}$ orbitals, and between neutron $\nu h_{11/2}$ and $d_{5/2}$ orbitals. Recent research has reported that three newly identified transitions populate the first $8_1^+$ state in $^{92}$Zr~\cite{PhysRevC.111.034312}.
\begin{figure}[ht]
    \vspace{2em}
    \centering
    \includegraphics[scale=0.47]{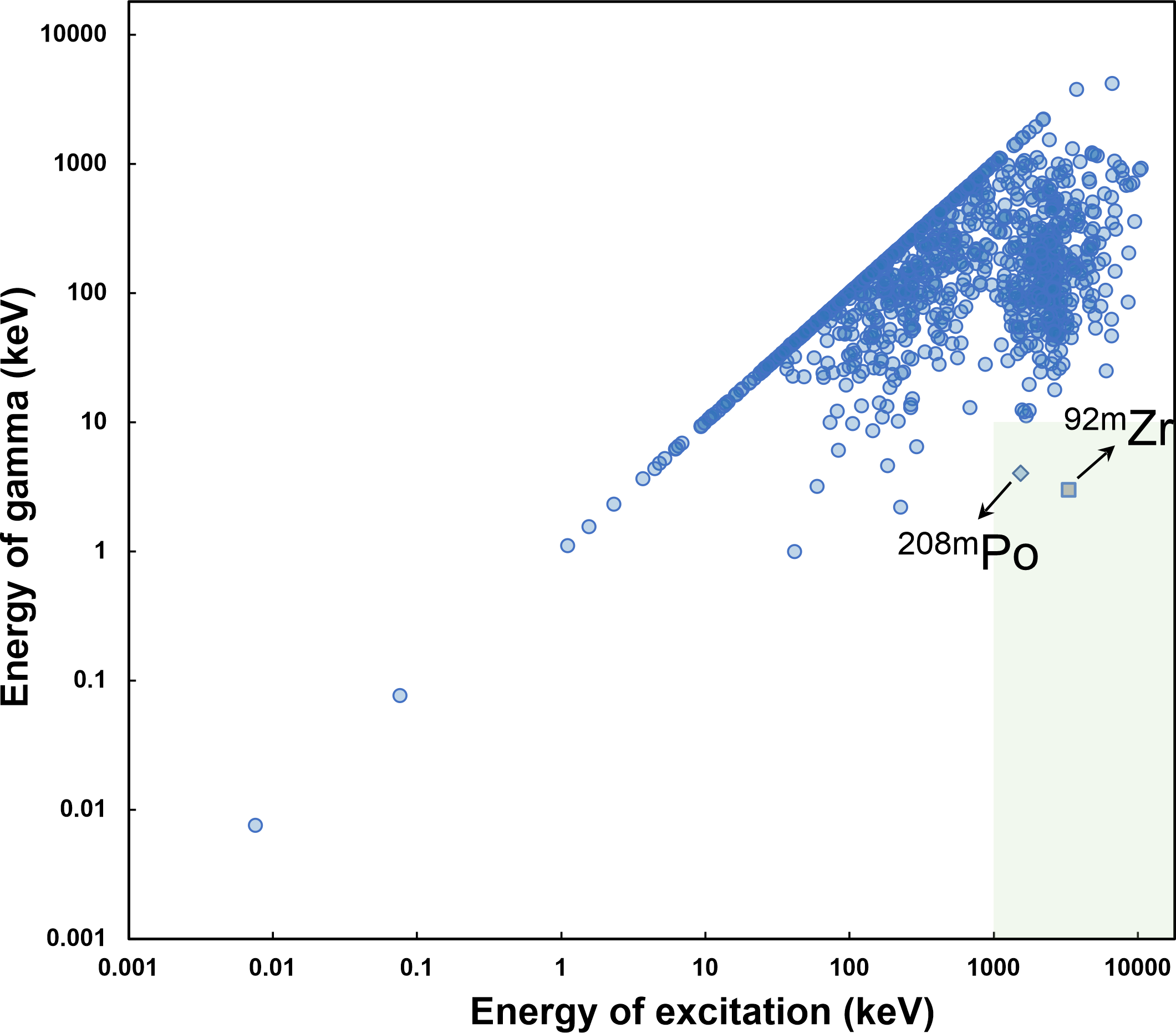}
    \caption{The scatter diagram correlates the excitation energy and depopulating $\gamma$-ray energy of isomeric nuclei with a single decay channel\cite{Garg_2023}.}
    \label{transition}
\end{figure}\par

Such isomers, especially those with excitation energy $\ge$ 1MeV that decay via $E_{\gamma} \le $ 10 keV transitions, are ideal candidates to satisfy the requirements of a four-level scheme as illustrated in Fig. \ref{4level}. As shown in Fig. 5, only two isomers, $^{92m}$Zr and $^{208m}$Po, meet the NGL conditions\cite{Garg_2023}. They can be produced as beams in a highly-charged atomic state using an accelerator technique. After purification, they can be transported to ion traps or storage rings within a timescale compatible with their prolonged lifetime. By injecting them into a material or irradiating them with an electron beam, the lifetime can be restored to a short value through the onset of internal conversion, whereas the upper level E$_2$ becomes highly populated. In such a case, the lower level E$_1$ is rarely occupied due to its very short lifetime, which results from a much larger energy of depopulating $\gamma$ transition. In $^{208}$Po, the lifetime of the E$_1$ level is approximately two orders of magnitude shorter than that of the E$_2$ level (85 ps for E$_1$ compared to 4.0 ns for E$_2$). Likewise, in $^{92}$Zr, the transition energy de-exciting the E$_1$ ($6^+$) state is four times that of the E$_0$ ($4^+$) state, leading to an estimated 3 orders of magnitude shorter lifetime for the E$_1$ level, assuming a similar B(E2) value. $^{92m}$Zr has an advantage here since its lifetime is probably comparable with that of the upper $8^+$ level. Compared with $^{208m}$Po, $^{92m}$Zr has a shorter neutral lifetime and can reach the dynamic equilibrium for stable laser emission more rapidly. 

Another difficulty for the $\gamma$ laser lies in the kinematic shift of $\gamma$ rays. In principle, the isomeric ions can be cooled to decrease the thermal movement and to increase the density. It can be manipulated to reach a specific relative velocity among different bunches of ions in order to offset the recoil energy loss using a Doppler shift. 

In conclusion, a cascade of transitions has been observed depopulating the first $8^+$ state in $^{92}$Zr at the detection station located far away from the primary fusion-evaporation reaction site. These $\gamma$ rays are inferred to originate from an unobserved state lying above the $8^+$ state. In neutral atoms, the lifetime of this state is only a few nanoseconds but is prolonged tremendously in highly charged states due to the blocking of the internal conversion decay process. Based on analysis of lifetimes during flight and after implantation, this state is proposed to have a spin and parity of 8$^-$ and decay via a very low energy E1 transition to the first 8$^+$ state. The decay scheme of this isomer constitutes a four-level structure that can be used to achieve population inversion of the $\gamma$ laser. Such high-lying isomer depopulating by low-energy transitions may provide a new approach for development of the $\gamma$ laser.\par

This work has been supported by the National Natural Science Foundation of China (Grants No. 12375128, No. 12121005, No. 12441506, No. 12135004, No. 12475129), the Major Science and Technology Projects in Gansu Province (Grant No. 24GD13GA005), the Open Project of Guangxi Key Laboratory of Nuclear Physics and Nuclear Technology (Grant No. NLK2023-08), the Central Government Guidance Funds for Local Scientific and Technological Development, China (Grant No. Guike ZY22096024), the computational resources from Sun Yat-sen University, the National Supercomputer Center in Guangzhou, the Research Program of State Key Laboratory of Heavy Ion Science and Technology Institute of Modern Physics Chinese Academy of Sciences (Grant No. HIST2025CS04), and the Guangdong Major Project of Basic and Applied Basic Research (Grant No. 2021B0301030006).\par
\bibliography{apssamp}

\end{document}